# Donors and Deep Acceptors in β-Ga$_2$O$_3$

Adam T. Neal[1,a], Shin Mou[1,a], Subrina Rafique[2], Hongping Zhao[2,3,4], Elaheh Ahmadi[5], James S. Speck[6], Kevin T. Stevens[7], John D. Blevins[8], Darren B. Thomson[8], Neil Moser[8], Kelson D. Chabak[8], Gregg H. Jessen[8]

[1] Air Force Research Laboratory, Materials and Manufacturing Directorate, Wright Patterson AFB, OH, 45433, USA
[2] Department of Electrical Engineering and Computer Science, Case Western Reserve University, Cleveland, OH, 44106, USA
[3] Department of Electrical and Computer Engineering, The Ohio State University, Columbus, OH 43210, USA
[4] Department of Materials Science and Engineering, The Ohio State University, Columbus, OH 43210, USA
[5] Department of Electrical Engineering and Computer Science, University of Michigan, Ann Arbor, MI, 48103, USA
[6] University of California, Santa Barbara, Santa Barbara, CA, 93106, USA
[7] Northrop Grumman SYNOPTICS, Charlotte, NC, 28273, USA
[8] Air Force Research Laboratory, Sensors Directorate, Wright Patterson AFB, OH, 45433 USA
[a] Electronic Address: shin.mou.1@us.af.mil and adam.neal.3@us.af.mil

**Abstract**

We have studied the properties of Si, Ge shallow donors and Fe, Mg deep acceptors in β-Ga$_2$O$_3$ through temperature dependent van der Pauw and Hall effect measurements of samples grown by a variety of methods, including edge-defined film-fed (EFG), Czochralski (CZ), molecular beam epitaxy (MBE), and low pressure chemical vapor deposition (LPCVD). Through simultaneous, self-consistent fitting of the temperature dependent carrier density and mobility, we are able to accurately estimate the donor energy of Si and Ge to be 30 meV in β-Ga$_2$O$_3$. Additionally, we show that our measured Hall effect data are consistent with Si and Ge acting as typical shallow donors, rather than shallow DX centers. High temperature Hall effect measurement of Fe doped β-Ga$_2$O$_3$ indicates that the material remains weakly n-type even with the Fe doping, with an acceptor energy of 860 meV relative to the *conduction band* for the Fe deep acceptor. Van der Pauw measurements of Mg doped Ga$_2$O$_3$ indicate an activation energy of 1.1 eV, as determined from the temperature dependent conductivity.

Excellent performance improvements in $Ga_2O_3$ power electronics transistors have been made since the first demonstrations of $Ga_2O_3$ MESFETs[1] and MOSFETs.[2] Breakdown voltages for $Ga_2O_3$ Schottky diodes have reached 1.1 kV [3] and 1.6 kV,[4] while breakdown voltages for MOSFETs are as high as 740 V.[5] Lateral device electric fields of at least 3.8 MV/cm [6] and vertical device electric fields of 5.1 MV/cm [3] have been demonstrated, along with on-currents of 1.5 A/mm.[7,8] Radio frequency operation of $Ga_2O_3$ MOSFETs, with $f_t$ and $f_{max}$ as high as 3.3 GHz and 12.9 GHz, respectively, has also been reported.[9] The rapid pace of $Ga_2O_3$ device development can be attributed in part to the effective and controllable n-type doping of $Ga_2O_3$, which has been achieved using tin (Sn),[1-17] silicon (Si),[17-24] and, more recently, germanium (Ge),[25] consistent with results from DFT calculations.[26] Some previous studies have examined the transport properties of shallow donors in $Ga_2O_3$; however, there remains some discrepancies regarding the energies of the shallow donors. Estimates of the donor energies range from 7.4 meV to 60 meV for Sn,[2,27,28] 16 meV to 50 meV for Si,[29-35] and we have previously reported a 17.5 meV donor energy for Ge.[36] Recent EPR studies report that Si may also exhibit a $DX^-$ state at energy 49 meV;[34] however, other groups have reported no evidence for a $DX^-$ state.[37] In addition to the shallow donor impurities, some impurities have been reported to induce insulating behavior in $Ga_2O_3$, acting as deep acceptors, including magnesium (Mg)[38-41] and iron (Fe).[42,43]

Given the wide range of donor energies reported in the literature for shallow donors and the limited data available on the deep acceptors, we have undertaken a study to understand the transport properties of Si, Ge, Fe, and Mg doped $Ga_2O_3$. In the case of shallow donors Si and Ge, by simultaneously and self-consistently fitting both the temperature dependent mobility and temperature dependent carrier density, we are able to accurately determine the donor energy for both Si and Ge to be 30 meV. Additionally, our transport measurements are consistent with Si and Ge acting as typical shallow donors, rather than shallow DX centers. We also report that EFG grown Fe doped $Ga_2O_3$ remains weakly n-type, as determined by high temperature Hall effect measurement, with the an acceptor energy for Fe of 0.86 eV relative to the $Ga_2O_3$ *conduction band*. Finally, we report that the conductivity of CZ grown $Ga_2O_3$ pulled from a melt with 0.1 mole percent MgO shows an activation energy of 1.1 eV, consistent with the previous report.[39]

To study the properties of shallow donors in $Ga_2O_3$, we have measured and analyzed the temperature dependent carrier density and mobility of several samples determined by van der Pauw and Hall effect measurements in a four terminal configuration. To accurately estimate the donor energies, the carrier density was fit using the charge neutrality equation[44] and the mobility fit using the solution to the Boltzmann transport equation in the relaxation time approximation,[45] including the Hall factor,[46] where ionized impurity,[47] neutral impurity,[48] and polar optical phonon[49] scattering mechanisms were included. An effective mass $m_* = 0.3 m_o$,[50-52] a low frequency relative dielectric constant $\kappa_S = 10$,[53,54] a high frequency relative dielectric constant of $\kappa_\infty = 3.5$,[54-56] an effective phonon energy $\hbar\omega = 44$ meV,[32] and an effective number of phonon modes $M = 1.5$ were used in the calculation. A detailed summary of the relevant equations from the cited references can be found in the supplementary material. Iteration was used to simultaneously and self-consistently fit both the carrier density and mobility. Uncertainty in the compensating acceptor concentration can propagate to the estimated donor energy when fitting the temperature dependent carrier density alone in moderately doped samples. With our approach, we are able to independently determine the compensating acceptor concentration from the ionized impurity limited mobility, allowing for more accurate estimation of the donor energies by avoiding said propagation. Others have used parts of this approach studying the properties of Si doped $Ga_2O_3$.[30-32]

The Si doped samples include a bulk CZ sample from Northrop Grumman Synoptics (Sample 3), a bulk EFG sample from Tamura Corporation (Sample 4), and two epitaxial films grown by LPCVD on c-sapphire substrates with 3.5° (Sample 5) and 6° (Sample 6) offcuts.[57] Glow discharge mass spectrometry (GDMS) and secondary ion mass spectrometry (SIMS) analysis of EFG $Ga_2O_3$ samples similar to Sample 4 and SIMS analysis of a CZ grown sample similar to Sample 3 confirm that Si is the dominant unintentional donor in the bulk melt-grown $Ga_2O_3$ used in this study. Our result is consistent with a previous study which determined that the unintentional Si doping comes from the $Ga_2O_3$ powder used as the source material for bulk melt-growth.[19] The LPCVD films were intentionally doped using $SiCl_4$. To make ohmic contacts, four 150 nm Ti/500 nm Au contacts were deposited on the sample edges and annealed in a tube furnace under Ar gas flow up to 450°C. Sample 1 and Sample 2 are Ge doped MBE grown $Ga_2O_3$ epitaxial films on semi-insulating substrates whose fabrication and growth details are published elsewhere.[25,36] Figure 1 and Figure 2 show the experimentally measured Hall carrier density and Hall mobility for the samples, along with the temperature dependent fittings. The mobility due to individual scattering mechanisms are shown for Sample 3 only. Parameters of the temperature dependent fittings for the samples are shown in Table I. The donor energies for the samples, along with several samples from the literature, are summarized in Figure 3. As the figure shows, the donor energies for samples seem to converge to a value of 30 meV as the donor concentration approaches $1 \times 10^{17}$ cm$^{-3}$ for both Si and Ge donors. However, as the donor concentration increases above $4 \times 10^{17}$ cm$^{-3}$, the donor energy begins to decrease, as is expected for highly doped semiconductors when an impurity band begins to form.[58,59] Consistent with this hypothesis, the decrease in the donor energy occurs as the donor density approaches $N_{dn} = (0.2/\alpha)^3 = 1.46 \times 10^{18}$ cm$^{-3}$, where $\alpha$ is the effective Bohr radius for gallium oxide, which is the estimated density at which a Mott metal-insulator transition would occur for the donor level.[58,59]

The above analysis assumes that the Si and Ge donors behave as typical shallow donors in Ga$_2$O$_3$, but recent EPR measurements have suggested that Si may behave as a shallow DX center.[34] By contrast, others have reported that they do not see evidence of DX center behavior in Ga$_2$O$_3$.[37] To address this open question in light of our Hall effect measurements, let us consider the charge neutrality equation for a DX center[60]

$$N_c \mathcal{F}_{1/2}\left(\frac{E_f - E_c}{kT}\right) + \frac{N_{dn}}{1 + \exp\left(\frac{(2E_{dn} + U) - 2E_f}{kT}\right) + 2\exp\left(\frac{E_{dn} + U - E_f}{kT}\right)} + N_{ac} = \frac{N_{dn}}{1 + \exp\left(\frac{-(2E_{dn} + U) + 2E_f}{kT}\right) + 2\exp\left(\frac{-E_{dn} + E_f}{kT}\right)} \quad (1)$$

where $N_c$ is the conduction band effective density of states, $N_{ac}$ the compensating acceptor concentration, $N_{dn}$ the donor concentration, $E_{dn}$ the donor energy, $\mathcal{F}_{1/2}$ the normalized Fermi-Dirac integral of order one half, $E_f$ the Fermi level, and $U$ the interaction energy for two electrons on a single donor. The interaction energy, $U$, is the energy difference between two electrons on a single donor (DX$^-$ state) and two non-interacting electrons on two different donors (neutral state). $U$ is negative, which indicates that the DX$^-$ state is the lower energy state, usually due to lattice distortion. For a normal shallow donor, $U$ is a large positive number due to coulomb repulsion of the electrons, and the conventional charge neutrality equation is recovered. Using Equation 1 and the parameters listed in the first two entries of Table II, a comparison of the temperature dependent carrier density and ionized impurity density for the typical normal shallow donor model and the shallow DX donor model are plotted in Figure 4. For the models in Figure 4, values for the normal donor model were chosen to be representative of the experimentally characterized samples listed in Table I, while the values for the DX donor model were chosen to match the temperature dependent carrier density of the normal donor model, with $U = -20$ meV as suggested by the recent EPR study.[34] As shown, it is possible to find a set of parameters for the DX donor model that almost match the temperature dependent carrier density of the normal donor model; however, as the dashed lines show, the density of ionized impurities at low temperatures is about a factor of five larger for the DX donor model. Because the density of ionized impurities is so much larger for the DX donor model, it underestimates the experimentally measured mobility of our samples at low temperatures where ionized impurity scattering dominates, as shown in Figure 5, with fitting parameters shown in Table II. This fact means that the DX donor model cannot be used to fit our measured Hall mobility and carrier density data. This analysis assumes that the spatial distribution of donors in the negatively charged $DX^-$ state and those in the typical positively charged ionized state are uncorrelated, so that all donors act as point-charge-like scattering centers. However, there is some evidence that the distributions of $DX^-$ donors and positively charged ionized donors can become correlated, acting as weaker dipole-like scattering centers.[61] While such a correlation may partially account for the discrepancy between our measured mobility data and the DX donor mobility models presented in Figure 5, it does not fully account for the factor of 4 to 5 discrepancy in low temperature mobility for Si doped Sample 3 and Sample 4. Therefore, our measured Hall data for these Si doped samples are consistent with a normal shallow donor and are inconsistent with a shallow DX center. This observation agrees with Irmscher et al.,[37] who also reports no evidence of DX behavior. We note, however, that we are only able to rule out DX center behavior in our samples because they have a compensation ratio $N_{ac}/N_{dn}$ less than one third. For compensation ratios of one third or greater in the normal donor model, it is always possible to find a set of parameters for the DX donor model that match both the temperature dependent carrier density and the low temperature ionized impurity density. This fact means that a normal shallow donor and shallow DX donor can only be distinguished using simultaneous, self-consistent carrier density and mobility fitting for compensation ratios less than one third, as is the case here.

Considering the remarkable consistency in experimental results across a wide variety of growth methods, including EFG, CZ, MBE, and LPCVD, along with the fact that Si and Ge are both group IV elements on the periodic table, we can conclude that Si and Ge act as typical shallow donors in all of the samples presented here.

In addition to measurements on Si and Ge donor doped samples, we have characterized Fe and Mg acceptor doped samples as well using the same four terminal van der Pauw and Hall effect measurement techniques. Due to the semi-insulating nature of these samples, high temperature Hall effect measurements were necessary to thermally activate carriers to enable Hall effect measurement. Measurements were performed under $N_2$ gas flow in a tube furnace with an external silicon carbide heater and electromagnet. Figure 6 shows the Hall carrier density for an Fe doped, EFG grown semi-insulating substrate from Tamura Corporation (Sample 7). Hall effect measurement was possible for temperatures above 400 K, with the sign of the Hall effect indicating that the β-$Ga_2O_3$ remains weakly n-type at elevated temperatures even with Fe doping. Glow discharge mass spectrometry analysis (GDMS) performed on a similar Fe doped sample indicates a concentration of about $1 \times 10^{17}$ cm$^{-3}$ for the unintentional Si donor and $8 \times 10^{17}$ cm$^{-3}$ for the intentional Fe acceptor. Therefore, we ascribe the observed n-type behavior to thermal activation of electrons on ionized Fe acceptors into the conduction band. Fitting with the charge neutrality equation yields an acceptor energy $E_c - E_{ac}$ of 860 meV. Note that this acceptor energy is referenced to the conduction band edge due to the n-type behavior. This acceptor energy near the conduction band is consistent with preliminary DFT calculations of Fe doped $Ga_2O_3$ which indicate that Fe induces midgap states closer to the conduction band.[43] The acceptor energy is somewhat higher than that recently observed via DLTS for Fe impurities in n-type bulk substrates;[62] however, this difference can be explained by the broadening of the Fe acceptor energy level at the much higher Fe concentration in this semi-insulating substrate. Because the carrier density does not saturate at higher temperatures, it is not possible to estimate the absolute value of the Fe and Si concentrations in this sample. However, the ratio of Fe acceptors to Si donors, $N_{ac}/N_{dn}$, is uniquely determined to be 1.65 by fitting the temperature dependent carrier density using the charge neutrality equation. We note that this ratio is smaller than one would expect based on the GDMS results, which could suggest that not all of the Fe dopants in the sample are electrically active. Finally, the inset of Figure 6 shows the temperature dependent conductivity measured by the van der Pauw method for an Mg doped sample grown by Northrop Grumman Synoptics using the CZ method (Sample 8). The melt from which the sample was pulled contained 0.1 mole % of MgO. Hall effect measurement was attempted, but it was not possible to resolve the Hall voltage due to low signal to noise ratio as a result of the very low conductivity for the sample. Least squares fitting of the temperature dependent conductivity indicates an empirical activation energy of 1.1 eV, which is consistent with a previous report on the activation energy of highly Mg doped $Ga_2O_3$.[39] Because Si contamination is present in these CZ grown samples, it is expected that the activation energy of 1.1 eV is approximately equal to the acceptor energy for Mg. However, without Hall effect measurement, it is not possible to determine the carrier type of the Mg doped sample. This fact means that we are unable to determine if this 1.1 eV activation energy is referenced to the conduction band or valence band edge of $Ga_2O_3$ from experiment, although recent DFT studies indicate the Mg acceptor level is closer to the valence band.[63]

In conclusion, simultaneous, self-consistent fitting of the temperature dependent carrier density and mobility of n-type β-$Ga_2O_3$, as measured by the van der Pauw and Hall effect methods, indicates a donor ionization energy of 30 meV for Si and Ge shallow donors. Accurate determination of the donor energy is enabled by reliable estimation of the compensating acceptor concentration through fitting of the low temperature ionzied impurity limited mobility. Additionally, comparsion of our Hall effect data to appropriate models indicates that Si and Ge act as typical shallow donors in the β-$Ga_2O_3$ samples presented here, as opposed to shallow DX centers. Finally, Fe doped β-$Ga_2O_3$ is shown to remain weakly n-type by high temperature Hall effect measurements, with an acceptor energy of 860 meV relative to the $Ga_2O_3$ *conduction band*.

**Supplementary Material**

See supplementary material for a summary of the relevant equations from the cited references used to fit the temperature dependent carrier density and mobility data.

**Acknowledgements**

The material is partially based upon the work supported by the Air Force Office of Scientific Research under award number FA9550-18RYCOR098. J.S.S. was also supported by the Defense Threat Reduction Agency through program HDTRA-17-1-0034. The content of the information does not necessarily reflect the position or the policy of the federal government, and no official endorsement should be inferred.

**Table I: Parameters of temperature dependent carrier density and mobility fitting for n-type β-Ga$_2$O$_3$ samples**

|  | Growth Method | Dopant | $N_{dn}$ $10^{16}$ cm$^{-3}$ | $E_{dn}$[a] meV | $N_{ac}$ $10^{16}$ cm$^{-3}$ | $N_{ac}/N_{dn}$ | $N_{neutral}$[b] $10^{16}$ cm$^{-3}$ | $\hbar\omega$[c] meV |
|---|---|---|---|---|---|---|---|---|
| Sample 1[d] | MBE | Ge | 6.5 | 28 | 4.8 | 0.74 | 80 | 44 |
| Sample 2 | MBE | Ge | 30 | 29 | 3.9 | 0.13 | 80 | 44 |
| Sample 3 | CZ | Si | 13 | 30 | 0.91 | 0.07 |  | 44 |
| Sample 4 | EFG | Si | 30 | 27 | 1.5 | 0.05 |  | 44 |
| Sample 5 | LPCVD | Si | 80 | 19 | 5.6 | 0.07 | 100 | 44 |
| Sample 6 | LPCVD | Si | 100 | 15 | 5.0 | 0.05 |  | 44 |
| Fornari et al. #12[e] | CZ | Si | 14.3 | 28.5 | 4.2 |  |  |  |
| Fornari et al. #3[e] | CZ | Si | 48.3 | 21.2 | 14 |  |  |  |
| Fornari et al. #7[e] | CZ | Si | 61.7 | 24.9 | 5.4 |  |  |  |
| Oishi et al.[f] | EFG | Si | 14 | 31 | 2.7 |  |  |  |

[a] Referenced to conduction band, [b] Additional neutral impurities beyond unionized donors
[c] Ref. 32, [d] A 2$^{nd}$ donor with $E_{dn2}$ =100 meV and $N_{dn2}$ = 1.5×10$^{16}$ cm$^{-3}$ was also included[33]
[e] Refs. 29,30 [f] Ref. 31

**Table II: Parameters for Comparing Normal Donor and DX Donor models**

|  | $N_{dn}$ $10^{16}$ cm$^{-3}$ | $E_{dn}$ meV | $U$ meV | $N_{ac}$ $10^{16}$ cm$^{-3}$ | $N_{ac}/N_{dn}$ |
|---|---|---|---|---|---|
| Normal Fig. 4 | 11 | 30 | $+\infty$ | 1.1 | 0.1 |
| DX Fig. 4 | 10 | 16 | $-20$ | 0 | 0 |
| Sample 2 DX, Fig. 5 | 27 | 16 | $-10$ | 5.4 | 0.2 |
| Sample 3 DX, Fig. 5 | 13 | 18 | $-10$ | 0 | 0 |
| Sample 4 DX, Fig. 5 | 25 | 10 | $-15$ | 0 | 0 |

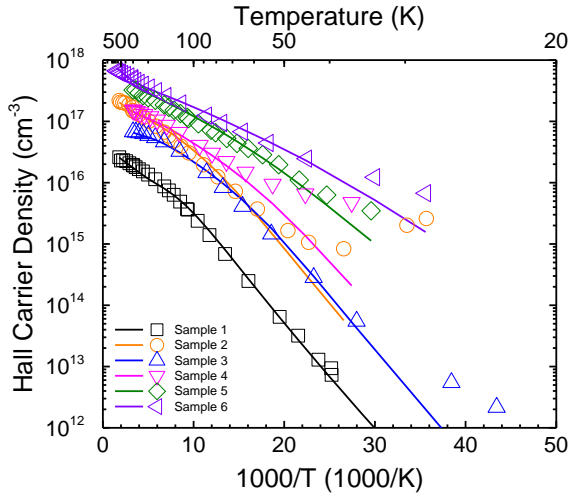

Figure 1: Measured Hall carrier density (symbols) and fittings (solid lines) for Si and Ge doped β-Ga$_2$O$_3$ samples.

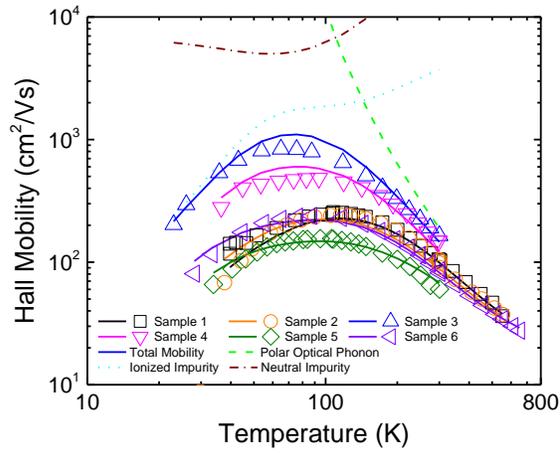

Figure 2: Measured Hall mobility (symbols) and fittings (solid lines) for Si and Ge doped β-Ga$_2$O$_3$ samples. Individual components of the mobility are shown for Sample 3, with the scattering mechanisms indicated in the legend.

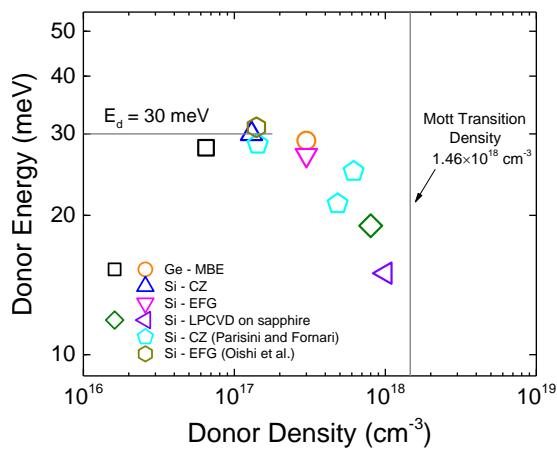

Figure 3: Summary of the donor energies as a function of donor concentration. Data from the literature [30,31] are also included as indicated in the figure legend.

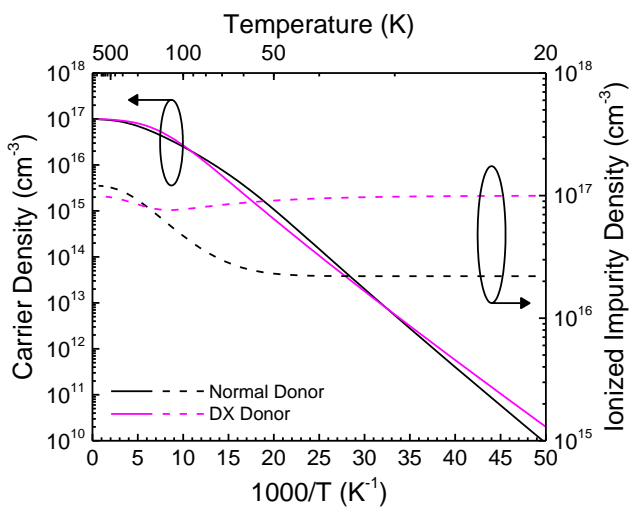

Figure 4: Comparison of carrier density and ionized impurity density vs. temperature for a normal donor and a DX center with $N_{dn} - N_{ac}$ of $1 \times 10^{17}$ cm$^{-3}$. Other parameters for the two models are shown in Table II. The higher concentration of ionized impurities at low temperature in the DX donor model leads to underestimation of the experimentally measured Hall mobility as shown in Figure 5.

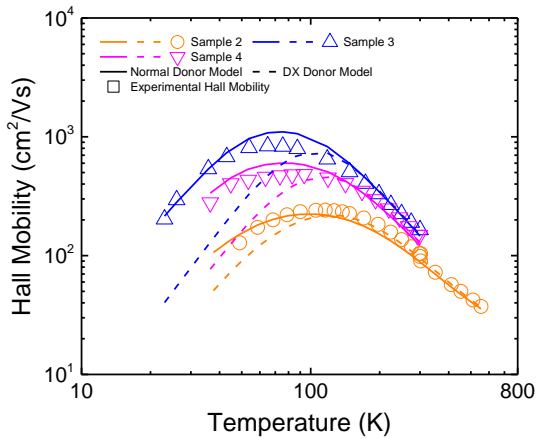

Figure 5: Comparison of mobility fitting using the normal donor model (solid lines) and the DX donor model (dashed lies) to the experimentally measured Hall mobility for Sample 2, Sample 3, and Sample 4. Only the normal donor model can fit the experimental data, as the DX donor model yields too much ionized impurity scattering at low temperatures and underestimates the experimentally measured mobility.

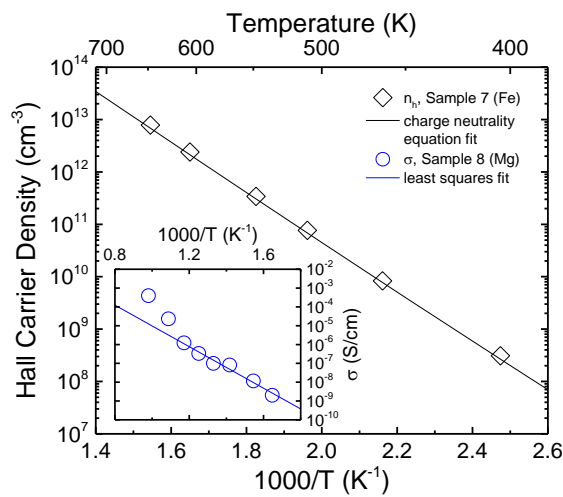

Figure 6: High temperature Hall carrier density for Sample 7, an Fe doped semi-insulating $\beta$-$Ga_2O_3$ sample. The sign of the Hall effect indicates that the sample is weakly n-type at elevated temperatures. Inset: Temperature dependent conductivity of Sample 8, an Mg doped semi-insulating $\beta$-$Ga_2O_3$ sample.

# Supplementary Material: Donors and Deep Acceptors in β-Ga$_2$O$_3$

Adam T. Neal[1,a)], Shin Mou[1,a)], Subrina Rafique[2], Hongping Zhao[2, 3, 4], Elaheh Ahmadi[5], James S. Speck[6], Kevin T. Stevens[7], John D. Blevins[8], Darren B. Thomson[8], Neil Moser[8], Kelson D. Chabak[8], Gregg H. Jessen[8]

[1] Air Force Research Laboratory, Materials and Manufacturing Directorate, Wright Patterson AFB, OH, 45433, USA
[2] Department of Electrical Engineering and Computer Science, Case Western Reserve University, Cleveland, OH, 44106, USA
[3] Department of Electrical and Computer Engineering, The Ohio State University, Columbus, OH 43210, USA
[4] Department of Materials Science and Engineering, The Ohio State University, Columbus, OH 43210, USA
[5] Department of Electrical Engineering and Computer Science, University of Michigan, Ann Arbor, MI, 48103, USA
[6] University of California, Santa Barbara, Santa Barbara, CA, 93106, USA
[7] Northrop Grumman SYNOPTICS, Charlotte, NC, 28273, USA
[8] Air Force Research Laboratory, Sensors Directorate, Wright Patterson AFB, OH , 45433 USA
a) Electronic Address: shin.mou.1@us.af.mil and adam.neal.3@us.af.mil

To fit the Hall data, first the temperature dependent carrier density is fit using the charge neutrality equation to make an initial guess at the concentrations and energies of the donors and acceptors in the material:[1]

$$N_c \, \mathcal{F}_{1/2}\left(\frac{E_f - E_c}{kT}\right) + \frac{N_{ac}}{1 + 2\exp\left(\frac{E_{ac} - E_f}{kT}\right)} = \frac{N_{dn}}{1 + 2\exp\left(\frac{E_f - E_{dn}}{kT}\right)} \qquad (S1)$$

$N_c$ is the conduction band effective density of states, $N_{ac}$ the acceptor concentration, $N_{dn}$ the donor concentration, $E_{ac}$ the acceptor energy, $E_{dn}$ the donor energy, and $\mathcal{F}_{1/2}$ the normalized Fermi-Dirac integral of order one half. $N_c$ is calculated analytically[1] using an electron effective mass of $0.3 m_o$.[2-4] With this initial fitting, we can then calculate the temperature dependent conduction band carrier density, the temperature dependent ionized impurity density, and the temperature dependent neutral impurity density. Next, those temperature dependent quantities are input into the appropriate models for ionized impurity scattering, neutral impurity scattering, and polar optical phonon scattering rates in Ga$_2$O$_3$. The scattering rate due to ionized impurities is:[5]

$$\frac{1}{\tau_{II}} = \frac{N_I q^4}{16\sqrt{2m_*}\pi\kappa_S^2\varepsilon_0^2}\left[\ln(1+\gamma^2) - \frac{\gamma^2}{1+\gamma^2}\right]E^{-3/2} \tag{S2}$$

$$\gamma^2 = \frac{8m_*EL_D^2}{\hbar^2} \tag{S3}$$

where $m_*$ is the electron effective mass, $\kappa_S$ the relative dielectric constant, and $L_D$ the Debye length due to screening of ionized impurities by conduction band free electrons. The neutral impurity scattering rate is:[6]

$$\frac{1}{\tau_{ni}} = \frac{(4\pi)(20)\kappa_S\varepsilon_0 N_{ni}\hbar^3}{m_*^2 q^2} \tag{S4}$$

And the polar optical phonon scattering rate is:[7]

$$\frac{1}{\tau_{POP}} = \frac{q^2\omega_o\left(\frac{\kappa_S}{\kappa_\infty}-1\right)}{4\pi\kappa_S\varepsilon_0\hbar\sqrt{2[E/m_*]}}\left[N_o\sqrt{1+\frac{\hbar\omega_o}{E}} + (N_o+1)\sqrt{1-\frac{\hbar\omega_o}{E}} - \frac{\hbar\omega_o N_o}{E}\sinh^{-1}\left(\frac{E}{\hbar\omega_o}\right)^{1/2}\right.$$
$$\left. + \frac{\hbar\omega_o(N_o+1)}{E}\sinh^{-1}\left(\frac{E}{\hbar\omega_o}-1\right)^{1/2}\right] \tag{S5}$$

$$N_o = \frac{M}{e^{\hbar\omega_o/kT}-1} \tag{S6}$$

where $\kappa_S = 10$ is the low frequency relative dielectric constant,[8,9] $\kappa_\infty = 3.5$ the high frequency relative dielectric constant,[9-11] $\hbar\omega_o = 44$ meV the effective phonon energy,[12] and $M = 1.5$ the effective number of phonon modes. The total scattering rate is the sum $\tau_m^{-1} = \tau_{ni}^{-1} + \tau_{POP}^{-1} + \tau_{II}^{-1}$. Using these scattering rates, the Hall mobility and Hall factor can be calculated as:[13,14]

$$\mu_h = r_h \frac{q\langle\langle\tau_m\rangle\rangle}{m_*} \tag{S7}$$

$$r_h = \frac{\langle\langle\tau_m^2\rangle\rangle}{\langle\langle\tau_m\rangle\rangle^2} \tag{S8}$$

$$\langle\langle\tau_m\rangle\rangle = \frac{\int_0^\infty E^{3/2}\tau_m(E)f(E)\,dE}{\int_0^\infty E^{3/2}f(E)\,dE} \tag{S9}$$

where $\langle\langle\tau_m\rangle\rangle$ is the average momentum relaxation time, averaged over energy as shown. Finally, with the Hall factor $r_h$, the Hall carrier density can be calculated as:

$$n_h = \frac{n}{r_h} \tag{S10}$$

and the temperature dependent carrier density can be fit while including the effect of the Hall factor. By iterating between fitting the temperature dependent Hall carrier density, fitting the temperature dependent mobility, and calculating the Hall factor, a simultaneous, self-consistent fit of the Hall effect data is achieved.

**Supplementary References**